# ON POSSIBILITY OF MEASUREMENT OF THE ELECTRON BEAM ENERGY USING ABSORPTION OF RADIATION BY ELECTRONS IN A MAGNETIC FIELD

R.Melikian

Yerevan Physics Institute, Alikhanyan Brothers St. 2, Yerevan 375036, Armenia

**Abstract**

The possibility of the precise measurement of the electron beam energy using absorption of radiation by electrons in a static and homogeneous magnetic field in a range up to a few hundred GeV energies, was considered in [1]. With the purpose of experimental checking of this method in a range of several tens MeV energies, the possibility of measurement of absolute energy of the electron beam energy with relative accuracy up to $10^{-4}$ is examined in details.

## I. Introduction

The possibility of the precise measurement of the electron beam energy by means of radiation absorption in a static and homogeneous magnetic field in a range up to a few hundred GeV energies, was proposed earlier in [1]. In this article, with the purpose of experimental checking of this method, the possibility of measurement of absolute energy of the electron beam energy with relative accuracy up to $10^{-4}$ in a range of several tens MeV is examined in details. We take into account the influences of the laser beam diffraction, the spread of electrons over energies and the length of formation of radiation absorption on the absorption process, which was not considered in [1].

We consider the quantum-mechanical approach of the photon absorption by electron in the static and uniform magnetic field $\vec{B}$ [2, 3] which allows us to find out the new additional aspects. In particular, it is found that the kinematical restrictions on the photon absorption process lead to interesting selection effects in angles of propagation of photons which can be absorbed by electrons. This circumstance is crucial for finding the energy of electrons.

The electron beam energy can be determined by applying the condition of absorption of the circularly polarized radiation by electrons in a static and homogeneous magnetic field. We choose the laser wavelength and the length of the magnet depending on the length of the photon absorption formation. The events of photons absorption can be established by means of the radiation detector, measuring changes of the intensity of the laser beam after its interaction with the electron beam in a magnetic field. We assume that the typical electron beam energy spread is approximately of the order $10^{-3}$ and we emphasize that with the accuracy $10^{-4}$ will be measured the disposition of center of the electron beam distribution over energies.

The essential advantage of the radiation absorption method consists in the opportunity of using semiconducting detectors with high spectral sensitivity and the high-speed response which are necessary for registration of radiation absorption and are currently produced by the industry.

## II. Principle of the method

**2.1 The condition of radiation absorption by electron in a static and homogeneous magnetic field and its dependence on electron energy.**

The considered method for measurement of energy of an electron beam is based on using the dependence of the condition of absorption of the circularly polarized radiation by electrons in the presence of a constant and homogeneous magnetic field on the energy of electrons (Fig.1). Acceleration of electrons with a high gradient of acceleration by means of consecutive multiphoton absorption in a constant magnetic field was discussed in many papers [4-12]. However, the feasibility of measurement of an electron beam energy by the radiation absorption method is much simpler than the acceleration of electrons by the same method, because for measurement of the electron beam energy it is enough to have absorption of one or several photons. The laser intensity, necessary for the acceleration of electrons, is much greater than the laser intensity, necessary for the measurement of electron beam energy. Moreover, for measurement of electron beam energy it is necessary to use a constant and homogeneous magnetic field, while for electron beam acceleration it is necessary to use a complicated magnetic field profile [11,12].

Acceleration of electrons in the electromagnetic wave of a microwave range and in the magnetic field, has been experimentally confirmed in [13-19]. The obtained results are in the good agreement with the theoretical predictions regarding validity of the radiation absorption condition and the electrons energy gain.

We consider the quantum mechanical treatment of the absorption of circularly polarized laser radiation by electrons in the static magnetic field $\vec{B}$, directed along the z–axis [2, 3]. We assume that electrons and laser beam are injected into a magnetic field under small $\varphi_{max} \ll 1$ and $\theta_{max} \ll 1$ angles with respect to the z-axis, accordingly (Fig.1).

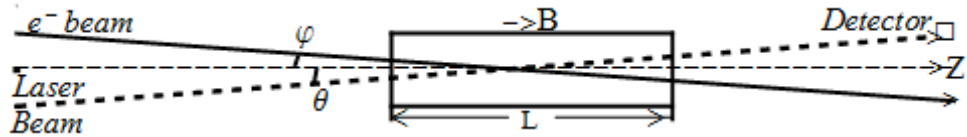

Fig.1. The schematic diagram of the set-up for measurement of electron beam energy.

The energy $\varepsilon_{P_z, n, \varsigma}$ of the electron in a magnetic field is determined by the well-known formula [20]:

$$\varepsilon_{P_z, n, \varsigma} = \left[ P_z^2 + m^2 + |e| B (2n + 1 + \varsigma) \right]^{1/2}, \quad (1)$$

where $n = 0, 1, 2,...$ labels the levels of the discrete spectrum of electron energy in a perpendicular direction to the magnetic field, $\varsigma = \pm 1$ is the projection of electron spin on a direction of $\vec{B}$, and $P_z$ is the z-component of the electron momentum. We use units for which $\hbar = c = 1$.

It is known, that in the constant and homogeneous magnetic field the component $P_{z,0}$ of the electron momentum is a continuous and conserved parameter [20]. After injection of electrons into a magnetic field they occupy some interval of energy levels $\varepsilon_{P_z, n, \varsigma}$ according to the spread of electrons velocities over angles $\varphi$. Absorption of a photon by an electron can occur at transition between the energies $\varepsilon_{P_z, n, \varsigma} \to \varepsilon_{P_z', n', \varsigma'}$, if the length $L$ of interaction of electron (Fig.1) with laser radiation with wavelength $\lambda$ in the presence of a magnetic field is greater than the length of formation of a photon absorption $l_a$. The length of formation of a photon absorption, as is known, is determined according to the formula [21-23]: $l_a \cong \lambda / (1 - \vec{\beta} \vec{k})$, where $1/(1 - \vec{\beta} \vec{k})$ is Doppler effect factor, $\vec{\beta} = \vec{V}/c$, $\vec{V}$ is the velocity of a electron, $\vec{k}$ is unit vector in a wave propagation direction. Thus, the necessary condition for the process of photon absorption by an electron has the form:

$$L \geq \ell_a \cong \frac{\lambda}{1 - \vec{\beta} \vec{k}} \quad (2)$$

Using the energy-momentum conservation for a photon absorption:

$$\varepsilon_{P_z,n,\varsigma} + \omega = \varepsilon_{P'_z,n',\varsigma}, \quad P_{z,0} + \omega\cos\theta = P_z \tag{3}$$

and expression (1), we find the photon absorption condition:

$$\varepsilon_{P_z,n,\varsigma} - Cos\theta\, P_{z,0} + \frac{\omega \sin\theta^2}{2} = \frac{\omega_c \nu\, m}{\omega}, \tag{4}$$

where $\varepsilon_{P_z,n,\varsigma}$ and $P_{z,0}$ are the initial energy and $z$-component of electron momentum, $\omega$ is the photon frequency, $\omega_c = eB/m$, $\nu \equiv n' - n = 1, 2, 3, \ldots$ . We take into account the transitions between the levels of electron energy without the change of electron spin direction, because the probability of spin-flip transitions is negligibly small [20], [21].

For absorption of radiation of optical and lower frequencies, and for $\theta \ll 1$, relevant to our case, we can neglect in (4) the quantum correction: $\frac{\hbar\omega}{mc^2}\frac{\sin\theta^2}{2} \lll \frac{\omega_c \nu}{\omega}$, and substituting into (4) the expression $P_{z,0}$ from (1), the photon absorption condition approximately becomes:

$$\omega = \frac{\omega_c (n' - n)}{\gamma\left(1 - Cos\theta \sqrt{1 - \frac{1}{\gamma^2} - \frac{2n\hbar\omega_c}{\gamma^2\, mc^2}}\right)} \tag{5}$$

where $\gamma = \varepsilon/mc^2$ is the electron relativistic factor.

The following conclusions can be drawn from the condition (5):

**a)** Neglecting the quantum correction $2n\hbar\omega_c/(\gamma^2 mc^2)$ and taking $n' - n = 1$ in expression (5), one obtains the condition of the classical radiation absorption [4], [8].

**b)** It is essential that the energy change $\Delta\varepsilon = \varepsilon_{P'_z,n'} - \varepsilon_{P_z,n}$ of the electron, due to absorption of a photon, depends on the angle $\theta$. In particular, for the case $Cos\theta = 0$, according to (5) we have:

$$\Delta\varepsilon = \frac{\omega_c (n' - n)}{\gamma}, \tag{6}$$

while in the relevant to us case $Cos\theta \neq 0$, $\theta \ll 1$ and $\gamma \gg 1$, we have:

$$\Delta\varepsilon \cong \frac{2\gamma \hbar\omega_c (n' - n)}{1 + \theta^2\gamma^2 + \dfrac{\hbar\omega_c}{mc^2} \cdot 2n}. \tag{7}$$

Comparing the relations (6) and (7) we see, that the change of electron energy due to photon absorption in the case of (7), is much greater. For example, for the parameters relevant to our case: $\gamma = 100$, $n' - n = 1$, $\omega_c = 0.83498 \cdot 10^{11}[\text{sec}^{-1}]$ (or $B \cong 4.7448 kGs$), $\theta = 3 \cdot 10^{-3}$, $n = 5 \cdot 10^7$ ($\varphi = 10^{-3}\, rad$) we have $\Delta\varepsilon \cong 0.97 \cdot 10^{-2}[\text{eV}]$. The value of $n$ can be estimated from the condition that in (1) the term $|e|B\,2n$ is approximately equal to the square of electron transverse momentum $P_\perp^2$, i.e., $P_\perp^2 = tg\varphi^2\, P_z^2 \cong |e|B\,2n \approx \varphi^2 \varepsilon^2$. Thus, the approximate value of $n$ can be determined from the expression:

$$\frac{\hbar\omega_c}{mc^2} \cdot 2n \cong \varphi^2\gamma^2. \tag{8}$$

**c)** From (5) and (7) it is clear that with increase of the quantum number n the value of $\Delta\varepsilon = \varepsilon_{P'_z,n'} - \varepsilon_{P_z,n}$ accordingly decreases, i.e. energy levels are not equidistant.

**d)** The change of energy $\Delta\varepsilon = \varepsilon_{P'_z,n'} - \varepsilon_{P_z,n}$ depends on energy of electron and below we shall consider the possibility of determination of electron energy using this dependence.

## 2.2 The length of a photon absorption formation.

It follows from (2) that, if the angle between directions of $\vec{V}$ and $\vec{k}$ vectors $\alpha \ll 1$ and $\gamma \gg 1$, then the length $L_M$ of the magnet, necessary for absorption of a photon by electron, will be determined by the condition:

$$L_M \geq \ell_a \cong \frac{2\lambda\gamma^2}{1+\alpha^2\gamma^2}. \qquad (9)$$

The dependence of length $\ell_a$ on the angle $\alpha$, for example, in case of $\gamma = 10^2$ and for $\lambda = 1.06 \cdot 10^{-3}\,[cm]$, according to the formula (9), is illustrated on Fig. 2.

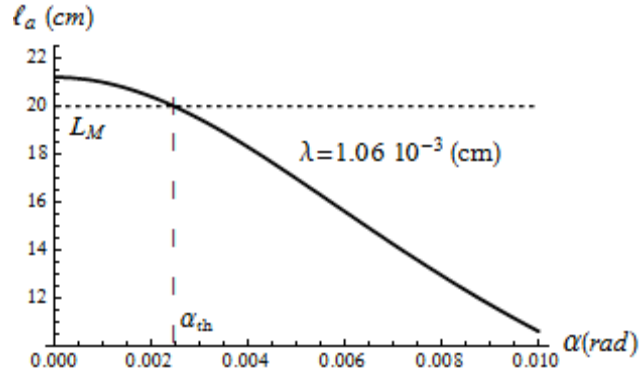

Fig. 2. The dependence of length a photon absorption formation $\ell_a$ on the angle $\alpha$. $\alpha_{th} \cong 2.45 \cdot 10^{-3}$ is the threshold angle, when $\ell_a = L_M$.

For absorption of a photon by electron it is essential that the time $t_{int} \cong \ell_a/c$ in which the photon can interact with the electron must be short compared with the lifetime $\tau = \hbar/\Gamma$ ($\Gamma$ - is the width of the energy level) of the electron on the given energy level, i.e. $\ell_a/c < \tau$. Using (9) we obtain:

$$\Gamma < \frac{\hbar\omega(1+\alpha^2\gamma^2)}{4\pi\gamma^2}, \qquad \frac{\Gamma}{\Delta\varepsilon} < \frac{1+\alpha^2\gamma^2}{4\pi\gamma^2}. \qquad (10)$$

For example, for the parameters relevant to our case: $\gamma = 100$, $\omega = 1.778 \cdot 10^{14}\,[sec^{-1}]$, $\alpha = 3 \cdot 10^{-3}$, we have: $\Gamma \cong 10^{-6}\,[eV]$ and $\Gamma/\Delta\varepsilon \cong 0.9 \cdot 10^{-5}$.

According to the principle of uncertainty the following condition must be satisfied [3, 20, 22]:

$$\Delta\varepsilon > \Gamma. \qquad (11)$$

From (10) it is clear, that for the parameters relevant to our case one, in fact, has $\Delta\varepsilon \gg \Gamma$, and the condition (11) is satisfied.

The photon absorption formation time $\tau$ is related by the principle of uncertainty with the frequency resolution of the incident wave $\Delta\omega$ in such a way as to satisfy the following requirement [3, 20, 21, 22]:

$$\Delta\omega \cdot \tau > 1. \qquad (12)$$

Taking into account the relation (10) we see that the criterion (12) will be satisfied, if one sets the restriction:

$$\gamma > \sqrt{\frac{1}{4\pi\left(\frac{\Delta\omega}{\omega}\right) - \alpha^2}}. \qquad (13)$$

For example, for the parameters interesting to us: $\Delta\omega/\omega = 10^{-4}$, $\alpha = 3 \cdot 10^{-3}$, we have: $\varepsilon \geq 14.5\,MeV$.

## 2.3 Determination of electron energy. The relative accuracy of electron beam energy.

From (5) we obtain for the $\gamma$ - factor of an electron:

$$\gamma = \frac{1}{\sin\theta^2}\left[\frac{v\omega_c}{\omega} \pm \cos\theta\sqrt{\left(\frac{v\omega_c}{\omega}\right)^2 - \left(1 + \frac{\hbar\omega_c\,2n}{mc^2}\right)\cdot\sin\theta^2}\right]. \tag{14}$$

The dependence of the electron energy $\varepsilon$ on the angle $\theta$, according to the formula (14), for example, for the concrete values of $v$, $\omega_c$, $\lambda$ and $n_{max}$, is illustrated on Fig.3.

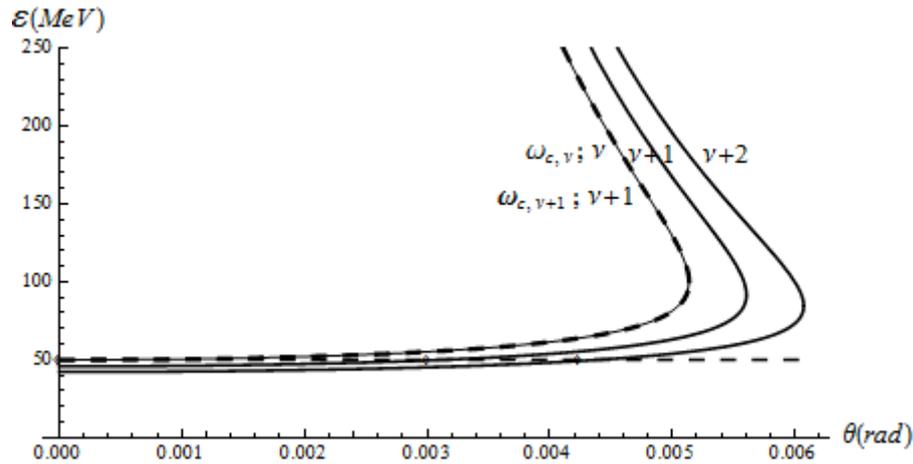

Fig. 3. The dependence of electron energy $\varepsilon$ on angle $\theta$. Parameters used here are: $n_{max} = 5\cdot 10^7$, $\omega_{c,v} = 0.83498\cdot 10^{11}\,[\sec^{-1}]$ ($B \cong 4.7448\,kGs$), $\omega_{c,v+1} = 0.76471\cdot 10^{11}\,[\sec^{-1}]$ ($B \cong 4.3455\,kGs$), $\lambda = 1.06\cdot 10^{-3}\,cm$, $v = 11$.

From the relations (5), (14) it is clear, that:
**a)** The energy of the electron reaches minimum at the angle $\theta = 0$ (Fig. 3) and is equal to

$$\gamma_{min} = \frac{\omega}{2\omega_c v} + \frac{\omega_c v}{2\omega} + \frac{\hbar\omega}{mc^2}\frac{n}{v}. \tag{15}$$

**b)** The value of $v$ should satisfy the restriction:

$$v \geq \frac{\omega\sin\theta}{\omega_c}\sqrt{1 + \frac{\hbar\omega_c\,2n}{mc^2}}. \tag{16}$$

From (14)-(16) it follows that one can use for absorption of a photon both at the transitions $v = 1$ and at the transitions $v = 2,3,...$. Transitions with $v = 2,3,...$ allow to use weaker magnetic fields which is crucial from the practical point of view.

From (14) and Fig. 3 it is clear, when $\omega$ is constant, the curve $\varepsilon(\theta, v, \omega_{c,v})$ described by the parameters $v, \omega_{c,v}$ and the curve $\varepsilon(\theta, v+1, \omega_{c,v+1})$ described by the parameters $v+1, \omega_{c,v+1}$ can coincide, if the value of $\omega_{c,v+1}$ is chosen properly, i.e.

$$\varepsilon(\theta, v, \omega_{c,v}) = \varepsilon(\theta, v+1, \omega_{c,v+1}). \tag{17}$$

Writing down the formula (15) for $\gamma(v, \omega_{c,v})$ and $\gamma(v+1, \omega_{c,v+1})$ and using the relation (17) for the case $\theta = 0$, we can exclude from this system of equalities the member $\frac{\hbar\omega}{mc^2}\cdot n$. As a result, we obtain the following expression for $\gamma_{min}$:

$$\gamma_{min} = \gamma(v, \omega_{c,v}) = \gamma(v+1, \omega_{c,v+1}) = \frac{\omega}{2}\left(\frac{1}{\omega_{c,v+1}} - \frac{1}{\omega_{c,v}}\right) + \frac{1}{2\omega}\left[(v+1)^2 \omega_{c,v+1} - v^2 \omega_{c,v}\right]. \quad (18)$$

From Fig. 3 it is clear that for fixed parameters $\omega$, $\gamma$ and $\omega_c$ only the photons propagating under the certain angles $\theta$ can be absorbed by electrons depending on value of $v$. To find energy of electron beam it is necessary to find these angles $\theta$.

Since for the parameters interesting to us we have $\frac{\hbar \omega_c}{mc^2} \cdot \frac{2n}{\gamma^2} \ll 10^{-4}$ then, neglecting this term in (5) we shall approximately obtain:

$$\theta \cong ArcCos\left[\frac{\gamma}{\sqrt{\gamma^2-1}}\left(1 - \frac{v\omega_{c,v}}{\gamma\omega}\right)\right]. \quad (19)$$

The upper limit of the angle $\theta$ can be found from the relation (14):

$$\theta_{max} \leq \frac{v\omega_{c,v}}{\omega}. \quad (20)$$

At the same time, for the given parameters $\gamma$, $\omega$ and $\omega_c$ from (19) it follows (due to $Cos\theta \leq 1$) that, the quantity $v$ is limited by the condition:

$$v_{min} \geq \frac{\omega}{2\gamma\omega_{c,v}}. \quad (21)$$

The dependence of the angle $\theta$ on $\omega_c$ for the constant $\omega$, $\gamma$ and for various $v$ according to (19) is shown on Fig.4.

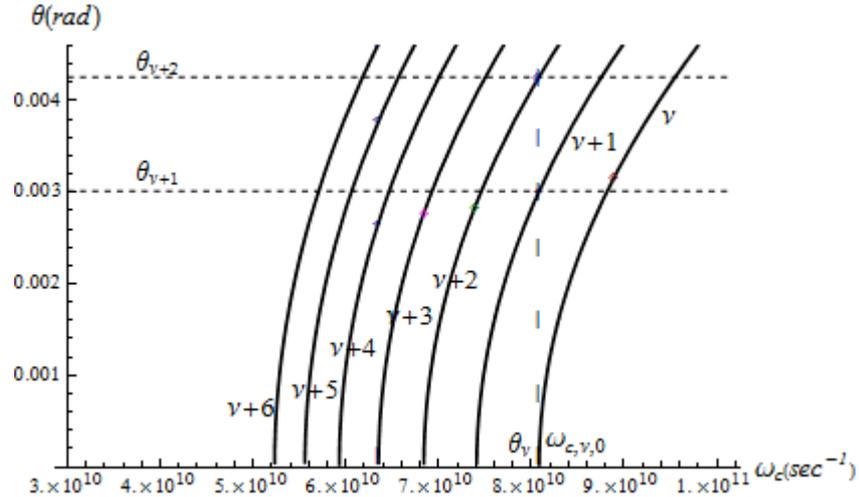

Fig.4. The dependence of the angle $\theta$ on $\omega_c$ for the constant $\omega$, $\gamma$ and in case of various values of $v$. Parameters used here are: $\omega_{c,v,0} = 8.0832 \cdot 10^{11}\,[sec^{-1}]$, $v = 11$, $\lambda = 1.06 \cdot 10^{-3}\,[cm]$, $\gamma = 10^2$. $\theta_{v+1} = 0.003015\,(rad)$, $\theta_{v+2} = 0.00426\,(rad)$.

From Fig.4 it is clear, that in case of some $\omega_{c,v,0}$ only the photons, propagating under angles $\theta_v = 0$, $\theta_{v+1}$ and $\theta_{v+2}$ can be absorbed by electrons. The angles of $\theta_{v+1}$ and $\theta_{v+2}$ can be found from the relation (19):

$$\theta_{v+1} = ArcCos\left[1 - \frac{\omega_{c,v,0}}{\omega\gamma\beta}\right] = ArcCos\left[1 - \frac{1-\beta}{v\beta}\right], \quad \theta_{v+2} = ArcCos\left[1 - \frac{2(1-\beta)}{v\beta}\right]. \quad (22)$$

Considering the absorption of photons for $\omega_{c,v,0}$ and $\omega_{c,v+1,0}$, from (19) we obtain:

$$(1-\beta)\omega\gamma = v\,\omega_{c,v,0} = (v+1)\omega_{c,v+1,0}. \qquad (23)$$

Having determined experimentally $\omega_{c,v,0}$ and $\omega_{c,v+1,0}$, we can find from (23) the minimal value of $v_{min}$:

$$v_{min} = \frac{\omega_{c,v+1,0}}{\omega_{c,v,0} - \omega_{c,v+1,0}}. \qquad (24)$$

The dependence of electron energy $\varepsilon$ on the angle $\theta$ for various parameters $\omega_c$ and $v$ according to (6) is shown on Fig.5. We assune that the electron beam has some spread over energy $\varepsilon$, i.e., absorption of photons by electrons is possible only in the interval of electron energy $\varepsilon - \Delta\varepsilon \le \varepsilon \le \varepsilon + \Delta\varepsilon$ (Fig.5a). For concreteness we shall assume that the electron beam has a Gaussian energy distribution (Fig.5b) with typical spread over energies $\Delta\varepsilon/\varepsilon = 10^{-3}$. Indeed, if $\gamma$ and $\omega$ are given, then according to (15) the value of $v$ is limited by the choice of reasonable quantity of $\omega_c$.

For the values $\omega_c \le \omega_{c,v,m}$ (in the case of concrete number $v = 11$), such that the curve $\varepsilon(\theta)$ lies outside of the interval $\varepsilon - \Delta\varepsilon \le \varepsilon \le \varepsilon + \Delta\varepsilon$ where the electrons are absent (Fig.5), it is clear that the intensity of photon absorption by electrons will be $I_{abs} = 0$ (Fig.6).

With the increase of the value $\omega_c > \omega_{c,v,m}$ due to increase of electrons number absorbing photons, the intensity of photon absorption $I_{abs}$ grows. For a certain value $\omega_{c,v,b}$, the curve $\varepsilon(\theta)$ passes over the bottom of the energy strip $\varepsilon - \Delta\varepsilon \le \varepsilon \le \varepsilon + \Delta\varepsilon$ and all electrons of the beam can absorb photons (Fig.5). As a result, the absorption intensity reaches the maximum $I_{abs,max}$ (Fig.6).

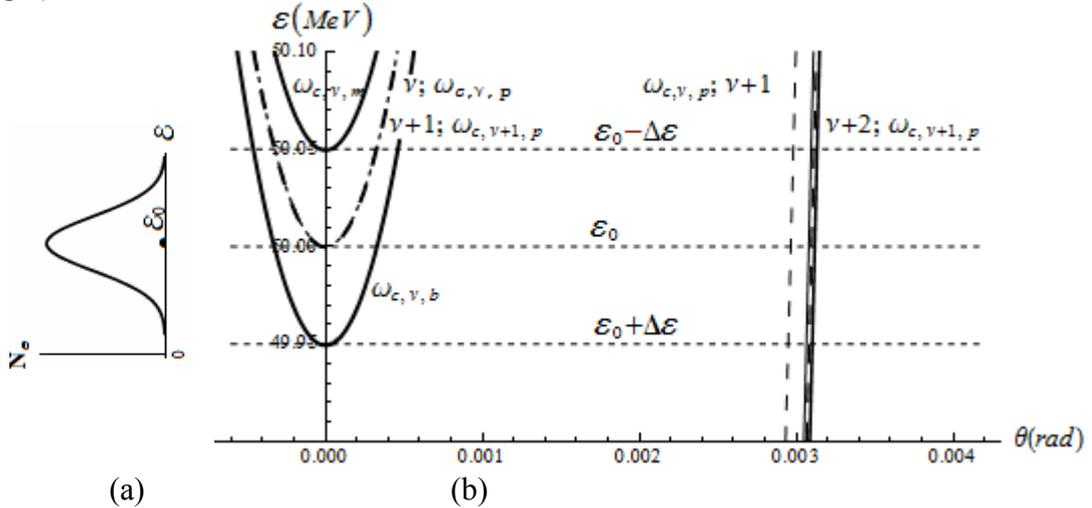

(a)      (b)

Fig.5. (a) Gaussian energy distribution of electrons. (b) The dependence of electron energy $\varepsilon$ on angle $\theta$ for various parameters $\omega_c$ and $v$. Parameters used here are: $v = 11$, $\omega_{c,v,m} = 0.83415 \cdot 10^{11} [\sec^{-1}]$, $\omega_{c,v,p} = 0.83498 \cdot 10^{11} [\sec^{-1}]$, $\omega_{c,v,b} = 0.83584 \cdot 10^{11} [\sec^{-1}]$ $\omega_{c,v+1,p} = 0.76472 \cdot 10^{11} [\sec^{-1}]$, $\lambda = 1.06 \cdot 10^{-3} [cm]$.

From Fig.5 it is clear that for a certain value of $\omega_{c,v,p}$ the curve $\varepsilon(\theta)$ passes through the midpoint of the energy interval $\varepsilon - \Delta\varepsilon \le \varepsilon \le \varepsilon + \Delta\varepsilon$ and only half of full number of electrons $N_e/2$ with energies $\varepsilon_0 \le \varepsilon \le \varepsilon_0 + \Delta\varepsilon$ will absorb photons Fig.6.

By measuring the photon absorption intensity $I_{abs,max}/2$ in case of $v$ and $v+1$, we can find the values of $\omega_{c,v,p}$ and $\omega_{c,v+1,p}$ (Fig.6), and, consequently, taking into account the relation (23), the electron beam energy $\varepsilon_0$ (Fig.5) can be calculated according to expression (18):

$$\varepsilon_0 = mc^2 \left( \frac{\omega}{2v\,\omega_{c,v,p}} + \frac{v\,\omega_{c,v,p}}{2\omega} \right). \tag{25}$$

Thus, according to (25) $\varepsilon_0$ can be found using the parameters $\omega$, $v$, $\omega_{c,v,p}$.

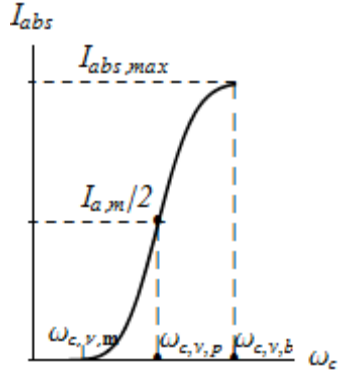

Fig.6. The dependence of photon absorption $I_{abs}$ by electrons on $\omega_c$. Parameters used here are: $\lambda = 1.06 \cdot 10^{-3}\,[cm]$, $v = 11$, $\omega_{c,v,m} = 0.83415 \cdot 10^{11}\,[\sec^{-1}]$, $\omega_{c,v,p} = 0.83498 \cdot 10^{11}\,[\sec^{-1}]$, $\omega_{c,v,b} = 0.83584 \cdot 10^{11}\,[\sec^{-1}]$.

The relative accuracy of electron beam energy can be determined from (25) by the following approximate formula:

$$\frac{\delta \varepsilon}{\varepsilon} \cong \frac{\delta \omega}{\omega} - \frac{\delta \omega_{c,v,p}}{\omega_{c,v,p}} \cong \frac{\delta \omega}{\omega} - \frac{\delta B}{B}. \tag{26}$$

Thus, the relative accuracy of electron beam energy depends on accuracy of measurement of laser frequency $\omega$ and on accuracy of measurement of magnetic field $B$. Consequently, the dependence $\varepsilon(\theta)$ on the diagram Fig. 5 will be represented not as a line, but as some strip with the width $\delta \varepsilon \cong \varepsilon_0 \cdot (\delta\omega/\omega - \delta B/B)$ according to (26).

The value of $\omega_{c,v,p}$ we can determine (Fig. 5, Fig. 6), if we take into account the spread of the electron beam energy. According to (19), the dependence of angle $\theta$ on $\omega_{c,v}$ for the constant $\omega$, $\gamma$ and in case of various values of $v$, taking into account the spread of the electron beam energy $\delta\varepsilon$, is shown on Fig.7. The intervals of angles $\delta\theta_v$, $\delta\theta_{v+1}$, $\delta\theta_{v+2}$ for the $\omega_{c,v,p}$ (Fig.7), which contribute to absorption of photons by electrons, can be found according to the formulas (22). The numerical estimations for parameters of Fig.7 show that: $\delta\theta_{v+2} < \delta\theta_{v+1} \ll \delta\theta_v$.

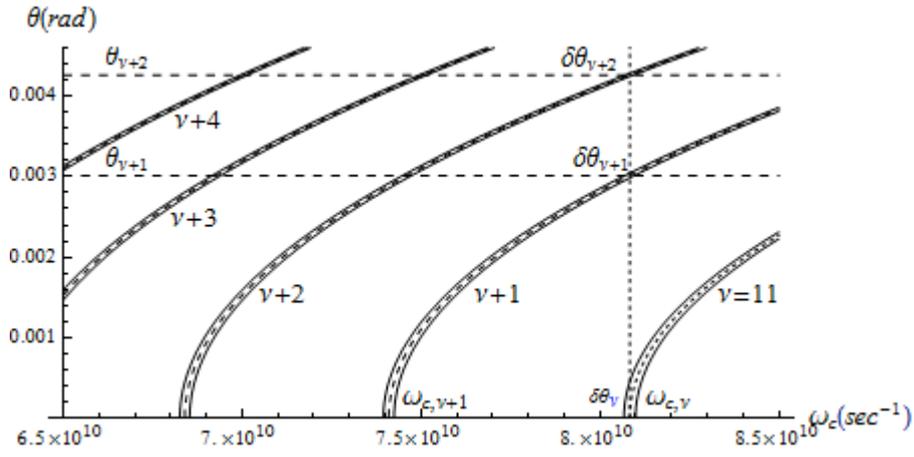

Fig.7. The dependence of angle $\theta$ from $\omega_{c,\nu}$ for the constant $\omega$, $\gamma$ and in case of various values of $\nu$. Parameters used here are: $\gamma = 10^2$, $\lambda = 1.06 \cdot 10^{-3}\,[cm]$, $\omega_{c,\nu,p} = 0.80832 \cdot 10^{11}\,[\sec^{-1}]$, $\omega_{c,\nu+1,p} = 0.7415 \cdot 10^{11}\,[\sec^{-1}]$, $\nu = 11$.

Moreover, because of diffraction of the light (see below (27), Fig.8) the intensity of light is maximal in the direction of $\theta = 0$ and $I(\theta_\nu) > I(\theta_{\nu+1}) > I(\theta_{\nu+2})$. Thus, we come to the conclusion, that for determining $\omega_{c,\nu,p}$ the dominating contribution to photon absorption by electrons is given by the photons, propagating in the interval of angles $\delta\theta_\nu$.

From (25) and Fig.5, Fig.6 it is clear that for finding of $\varepsilon_0$, it is necessary to tune the value of $B$ in the case of $\omega = const$ or to tune the value of $\omega$ in the case of $B = const$, depending on practical expediency.

It is obvious that measurement of $\varepsilon_0$ and $\varepsilon_0 - \Delta\varepsilon$ by the method described above (Fig.5) allows to find the real spread of electron beam $\Delta\varepsilon$.

## III. Influence of light diffraction on absorption of photons

It is known that the light beam of diameter D diverges because of diffraction in the range of angles $0 \leq \theta \leq \theta_d \cong \lambda/D$ around of direction of wave vector $\vec{k}$ of the incident wave. Distribution of the light intensity depending on the diffraction angle $\theta$ is determined by expression [22]:

$$I(\theta) = I_0 \left[\frac{2 J_1(\psi)}{\psi}\right]^2 \qquad (27)$$

and has the shape schematically represented on Fig. 8. Here $\psi = Dk\theta/2$, $k = \omega/c$, $J_1(\psi)$ is the cylindrical Bessel function of the first order, $I_0$ is the intensity of light propagating in the direction of $\theta = 0$. Let us note, that the light frequency does not vary due to diffraction.

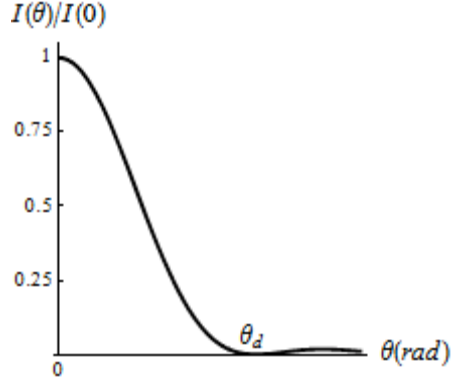

Fig.8. Distribution of light intensity depending on diffraction angle $\theta$.

From the restriction (20) it follows, that at any case: $\theta < \dfrac{\nu \omega_{c,\nu}}{\omega}$. The meaning of this restriction is clear from Fig.2, Fig.5 and Fig.7. Only the photons in the limited interval of angles $0 \le \theta \le \theta_f$ can take part in the process of absorption, i.e. a specific "collimation" of the light beam occurs. For the parameters of the electron and the laser beam interesting to us, usually we have (Fig.8): $\theta_f \ll \theta_d$. It is clear that the photons can only be absorbed if they propagate under angles smaller than $\theta_d$, i.e. $\theta_{\nu+1} < \theta_d$ and $\theta_{\nu+2} < \theta_d$. From Fig.7 and Fig.8 also it is clear, that: $I(\theta_\nu) \gg I(\theta_{\nu+1}) \gg I(\theta_{\nu+2})$, which is crucial for measurement of $\omega_{c,\nu,p}$.
The influence of the light diffraction on detection of photon absorption will be considered in section IX.

## IV. Estimation of the laser intensity required for absorption of photons

It is known that in a field of circularly polarized electromagnetic wave and in a magnetic field, the electrons can be accelerated due to absorption of laser photons [4, 3, 7, 9-11, 24]. Intensity of the laser required for RA of photons by an electron can be found using the classical formula for the growth of electron energy [4, 6, 8, 10-12] and by using the formulas (1), (4):

$$\Delta \gamma \cong \omega \xi \ell_a \beta_\perp \cong \omega \xi \ell_a \sqrt{\dfrac{2\nu \omega_c}{\omega \gamma_0}} \; . \qquad (28)$$

Here $\xi = eE/mc\omega$ is the parameter of the laser intensity, $E$ is the amplitude of electric field of an electromagnetic wave, $\ell_a$ is the length of formation of a photon absorption of energy $\Delta \varepsilon = \hbar \omega$ during time $t_a = \ell_a/c$.
If on the length of magnet $L_M > \ell_a$ the electron absorbs a photon with energy $\Delta \varepsilon = \hbar \omega$, then the relation (28) can be written in the form:

$$\hbar \omega \cong 19.4 \ell_a \sqrt{I_{las}} \sqrt{\dfrac{2\nu \omega_c}{\omega \gamma_0}} \; . \qquad (29)$$

Here we took into account the known relation between laser intensity $I_{las,a}$ and amplitude of electric field $E$ of electromagnetic wave:

$$E\left[\dfrac{V}{cm}\right] = 19.4 \sqrt{I_{las}\left[\dfrac{W}{cm^2}\right]}. \qquad (30)$$

Using (29), (25) and (9) for intensity of the laser required for absorption of a photon by electron, we obtain the following approximate value:

$$I_{las,a} \cong \left( \frac{\hbar\omega}{19.4} \frac{\omega}{4\pi c \gamma_0} \right)^2. \tag{31}$$

The dependence of laser intensity $I_{las,a}$ on energy of electrons $\varepsilon$ in case of $CO_2$ and $Nd:YAG$ lasers according to the formula (31), is illustrated in Fig. 9.

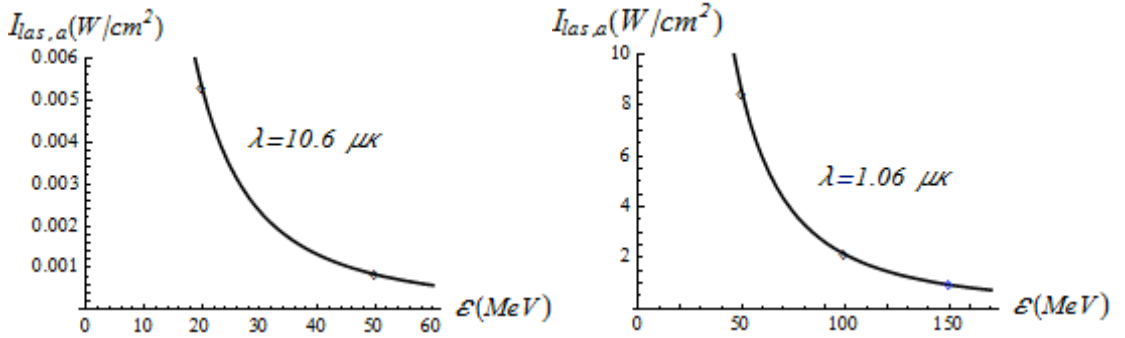

Fig.9. The dependence of laser intensity $I_{las,a}$ on energy of electrons $\varepsilon$ in the case of $CO_2$ and $Nd:YAG$ lasers.

## V. Choice of the magnet length taking into account the edge effects

For the choice of the magnet length $L_M$, we use the formulas (9) and (29) for the growth of the electron energy.
Since only the integer number of photons can be absorbed by electrons on the lengths of absorption $\ell_a, 2\ell_a,\ldots$ then according to (29), the magnet length $L_M$ can be chosen within the limits (Fig. 10):

$$\ell_a < L_M < 2\ell_a. \tag{32}$$

The condition (32) has a practical meaning and allows one to use the homogeneous part of the magnetic field, as well as to exclude the influence of edge effects.

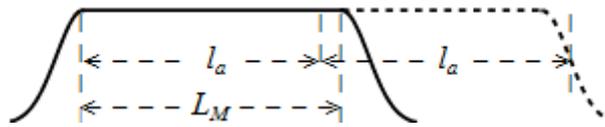

Fig.10. The choice of the magnet length $L_M$ within the limits of $\ell_a < L_M < 2\ell_a$ allows to exclude the influence of the edge effects.

From (9) it follows, that for electrons of high energy when $\alpha^2\gamma^2 \gg 1$ the restriction on the magnet length will be: $L_M \geq \ell_a \cong 2\lambda/\alpha^2$, i.e., in case of suitable choice of $\lambda$ and $\alpha$ the length of a magnet can be in reasonable limits. For example if $\lambda = 1\mu k$ and $\alpha = 3\cdot 10^{-3}$ then: $\ell_a \cong 23.6 cm$.

## VI. Exception of influence of accompanying radiation of electrons

The radiation of a photon by an electron in the considered fields can occur, if the length of interaction of the electron with the laser photons in the presence of a magnetic field is greater than the length of formation of a photon $l_r$.
The length of radiation formation, as is known, is determined according to the formula [21-23]:

$l_r \cong \lambda_r / (1 - \vec{\beta} \vec{k}_r)$, where $1/(1 - \vec{\beta} \vec{k}_r)$ is Doppler effect factor, $\vec{k}_r$ is a unit vector in the radiation direction, $\lambda_r$ is the length of radiation. Thus, the necessary condition for the radiation of a photon by an electron has the form:

$$L_M \geq \ell_r \cong \frac{\lambda_r}{1 - \vec{\beta} \vec{k}_r}. \tag{33}$$

It is known that in the case of $\gamma \gg 1$, the electron radiates basically in the direction of the movement in interval of angles $-1/\gamma \leq \alpha \leq 1/\gamma$ around the direction of electron velocity $\vec{V}$.
Thus, according to (29) the length of electron radiation will be determined according to the expression:

$$\lambda_r \leq \frac{L_M (1 + \alpha_r^2 \gamma^2)}{2\gamma^2}, \tag{34}$$

where $\alpha_r$ is the angle between vectors $\vec{k}_r$ and $\vec{V}$.

From (34) it follows, that the length of radiation in a case of $\alpha_r = 0$ (i.e. in direction of $\vec{V}$) will be minimal: $\lambda_{r,\min} \cong L_M / 2\gamma^2$ and $\lambda_r(\alpha_r = 1/\gamma) \cong 2\lambda_{r,\min}$, if $\alpha_r = 1/\gamma$. For example, if $L_M = 20 cm$, $\gamma = 100$ then: $\lambda_{r,\min} \cong 10^{-3} cm$ $\lambda_r(\alpha_r = 1/\gamma) \cong 2 \cdot 10^{-3} cm$.

We have found above that the greatest contribution to absorption is given by the laser photons propagating in the interval of angles $2\delta\theta_\nu$ (Fig.7) around the $z$-axis (Fig.1). If detector $D$ (Fig.1) is located on distance $l_d$ from the magnet then absorption of photons may be measured in the area with the diameter: $d_D = l_d 2\delta\theta_\nu$. For example, if $l_d = 5m$, $2\delta\theta_\nu = 10^{-3}$ then $d_D = 0.5 cm$.

Thus, on a given region with area $\pi d_D^2 / 4$, in addition to the laser radiation, there is also radiation of an electron, with the wavelength $\lambda_r(\alpha_r)$ greater than $\lambda$. Choosing the incidence angle $\varphi$ of the electron beam (Fig.1) and the distance $l_d$, we can always provide the condition $\lambda_r(\alpha_r) > \lambda$. This restriction allows us to exclude the influence of radiation on the result of measurement, provided we choose the semi-conductor detector to be sensitive only to the spectrum $\leq \lambda$. Thus, for measurement of the electron beam energy we shall register only the change of intensity of laser photons.

## VII. Influence of absorption of laser photons on parameters of electron beam

The parameters of the electron beam will not be changed significantly due to the absorption of laser photons because of the following reasons:
a) According to (3) for parameters interesting to us we have $\Delta\varepsilon/\varepsilon = \hbar\omega/\varepsilon \ll 10^{-4}$, therefore the change of electron beam energy due to absorption of photons will be negligible.
b) Let us estimate the change of direction of electron velocity $\vec{V}_e$ due to the absorption of laser photons. According to (3): $\Delta P_z / P_z = \hbar\omega Cos\theta / \varepsilon \ll 10^{-4}$.
Taking into account that: $P_\perp \cong \sqrt{eB 2n}$, $n' = n + \nu$ and $\nu/n \ll 10^{-4}$ we have $\Delta P_\perp / P_\perp \cong \nu/2n \ll 10^{-4}$. It means the change of direction of electron velocity is negligible.

## VIII. Influence of electron synchrotron radiation on the absorption process

It is known, that the intensity of synchrotron radiation in field $\vec{B}$ is different from zero if $\vec{B}$ has a transverse component $B_\perp \cong B\varphi$ to electron velocity $\vec{V}_e$. In addition, it is known that synchrotron radiation is formed on a length $\Delta \ell$ of electron trajectory, on which vector $\vec{V}_e$ turns on angle $\theta \cong 1/\gamma$ [20, 21]. From the geometry of the electron movement it follows, that:

$$\Delta \ell = \theta R, \qquad (35)$$

where: $R = V_e/\omega_H \cong c/\omega_H$, $\omega_H = eB_\perp/mc\gamma = \varphi \omega_c/\gamma$. Then, the relation (31) can be rewritten as:

$$\Delta \ell \cong \frac{c}{\omega_c \varphi}. \qquad (36)$$

According to (36), for parameters interesting to us $\omega_c$, $L_M$ and $\varphi \ll 1$, we have $\Delta \ell \gg L_M$, i.e., the influence of synchrotron radiation of electrons on length of magnet $L_M$ is inessential.

## X. Detection of photon absorption

The fact of absorption of photons by electrons can be established by measuring the ratio of absorbed photons number $N_{abs,ph}$ to the total number $N_{tot,ph}$ of laser photons per interaction time $\tau_{int}$ of electron beam with a laser beam on the length $\ell_a$.

The number $N_{abs,ph}$ of the photons absorbed by electron beam of length $\ell_{eb}$ can be estimated, if we take into account the fact that each electron passes through interaction area $\ell_a$ (Fig.10) only once, absorbing one photon. If the number of the electrons in the beam is equal to $N_{eb}$ then $N_{abs,ph} = N_{eb}$, independently of the fact whether $\ell_{eb} > \ell_a$ or $\ell_{eb} < \ell_a$.

The laser intensity $I_{las,a}$ (or the value of $E$ in accordance with (30)), necessary for absorption of one photon by an electron on length $\ell_a$, can be found according to (31). Then, the number of photons $N_{tot,ph}$ of the laser beam falling on a surface $S$ of the detector during of time $\tau_{int}$, can be determined by the relation:

$$N_{tot,ph} = \frac{I_{las,a}}{\hbar \omega} S \tau_{int}. \qquad (37)$$

The parameters of the laser ($I_{las,a}$, $D$, $\omega$), the detector and the magnet ($B$, $L$) should be chosen so that the change of laser intensity due to resonance absorption of photons:

$$\eta = \frac{N_{abs,ph}}{N_{tot,ph}} = \frac{N_{e,b} \hbar \omega}{I_{las,a}} \frac{4c}{\pi D^2 \ell_a} \qquad (38)$$

could be registered by the detector.

From (38) it is clear that with decrease $D$ the quantity $\eta$ increases. On the other hand, however, decreasing of $D$ is limited for two reasons: a) Firstly, to provide interaction of electrons and photons, the light beam diameter should be greater than diameter of electron beam; b) Secondly, when decreasing $D$, the diffraction angle $\theta_d \cong \lambda/D$ of the light beam increases (Section III).

## X. Conclusion

The possibility of measurement of absolute energy of the electron beam using absorption of laser radiation by electrons in a static and homogeneous magnetic field, with a view of experimental checking of this method in a range of several tens MeV is considered in details. This method allows determining the absolute energy of an electron beam with relative accuracy up to $10^{-4}$.

With the accuracy $10^{-4}$ can be measured just the energy of distribution center of electrons. The method also allows finding the real spread of electron beam $\Delta\varepsilon$.

The influence of the laser beam diffraction, the spread of electrons over energies and of the length of formation of photon absorption on the absorption process is considered. It is found that the kinematical restrictions on the photon absorption process lead to interesting selection effects for the angles of propagation of photons which can be absorbed by electrons.

Detectors with high spectral sensitivity and the high-speed response, which are necessary for registration of the radiation absorption with wavelength $1 \div 12\,\mu k$, are not unique and are produced industrially. The parameters of the electron beam will not worsen during measurement of the energy, which allows continuous monitoring of the electron beam energy.